\documentclass[aps,prl,twocolumn,showpacs]{revtex4-1}
\usepackage{bm}
\usepackage{graphicx}
\usepackage{amsmath,amsthm,amssymb}
\usepackage{dsfont}
\usepackage{soul}

\usepackage{color}
\begin{document}

\title{Topological transitions in spin interferometers}

\author{Henri Saarikoski,$^1$
 J. Enrique V\'azquez-Lozano,$^2$ Jos\'e Pablo Baltan\'as,$^2$  Fumiya Nagasawa,$^3$ Junsaku Nitta,$^3$ and Diego Frustaglia$^2$}
\affiliation{
$^1$RIKEN Center for Emergent Matter Science (CEMS), Saitama 351-0198, Japan
}
\email{e-mail: henri.saarikoski@riken.jp}
\affiliation{
$^2$Departamento de F\'isica Aplicada II, Universidad de Sevilla, E-41012 Sevilla, Spain
}
\email{e-mail: frustaglia@us.es}
\affiliation{
$^3$Department of Materials Science, Tohoku University, Sendai 980-8579, Japan
}

\date{\today}


\begin {abstract}
We show that topological transitions in electronic spin
transport are feasible by a controlled manipulation of spin-guiding fields.
The transitions are determined by the
topology of the fields texture through an \emph{effective} Berry phase
(related to the winding parity of spin modes around poles in the Bloch
sphere), irrespective of the actual complexity of the nonadiabatic spin dynamics. This manifests as a distinct dislocation of the interference
pattern in the quantum conductance of mesoscopic loops. The phenomenon
is robust against disorder, and can be experimentally exploited to
determine the magnitude of inner spin-orbit fields.
\end{abstract}

\maketitle


In the early 1980s Berry showed that quantum states in a cyclic motion may acquire a phase component of geometric nature \cite{berry}.
This opened a door to a class of topological quantum phenomena in optical and material systems~\cite{BMKNZ03}.
With the development of quantum electronics in semiconducting nanostructures,
a possibility emerged to manipulate electronic quantum states via the control of spin geometric phases driven by magnetic field textures~\cite{loss}.
After several experimental attempts \cite{MHKWB98,YPS02,BKSN06,KTHSHDSBBM06,GLIERW07} indisputable signatures of spin geometric
phases in conducting electrons were found in 2012 \cite{nagasawa1}
in agreement with the theory~\cite{frustaglia}. This paved the way for the development of a topological spin engineering~\cite{nagasawa2}. 

\begin{figure}
\includegraphics[width=\columnwidth]{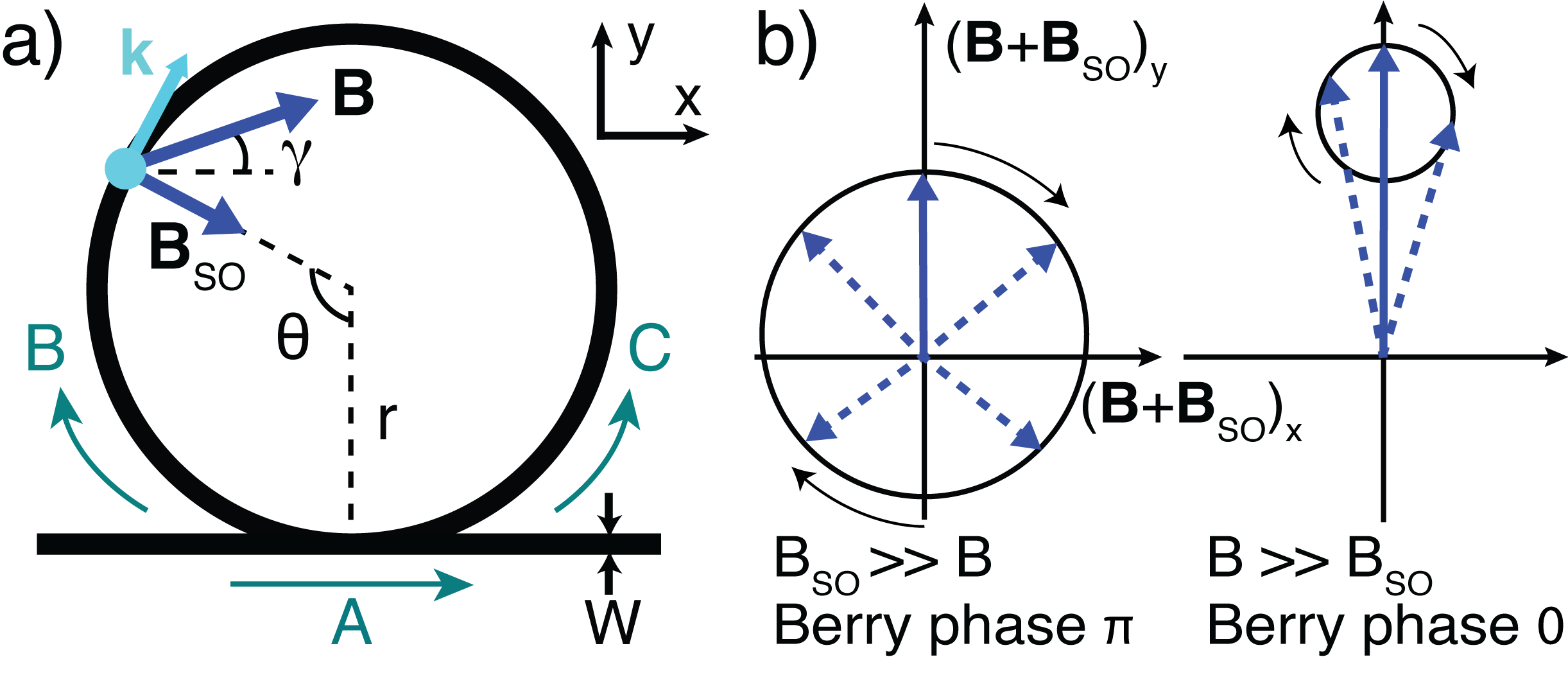}
\caption{a) The model system: a conducting wire of width $W$ is attached tangentially to a ring of radius $r$, forming a loop. The main interference paths are straight along the wire (A) and (counter)clockwise around the loop (B and C). The spin-orbit field ${\bf B}_{\rm SO}$ is radial and the homogeneous magnetic field ${\bf B}$ lies in the $xy$ plane. 
b) The Berry phases in the adiabatic limit. For $B_{\rm SO}\gg B$ the solid cone $\Omega=2\pi$ corresponding to the Berry phase $\pi$ (left).
For $B \gg B_{\rm SO}$ the solid cone vanishes giving Berry phase 0 (right).}
\label{fig-1}
\end{figure}

An early proposal for the topological manipulation of electron spins by Lyanda-Geller involved the abrupt switching of
Berry phases in spin interferometers~\cite{lyanda-geller}. These are conducting rings of mesoscopic size subject to Rashba spin-orbit
(SO) coupling, where a radial magnetic texture ${\bf B}_{\rm SO}$ steers the electronic spin (Fig.~\ref{fig-1}a). For relatively large field strengths
(or, alternatively, slow orbital motion) the electronic spins follow
the local field direction adiabatically during transport, acquiring a Berry phase factor $\pi$ of geometric origin (equal to half the solid angle
subtended by the spins in a roundtrip) leading to destructive interference effects. By introducing an additional in-plane
uniform field ${\bf B}$, it was assumed that the spin geometric phase undergoes a sharp transition at the critical point beyond which
the corresponding solid angle vanishes together with the Berry phase, and interference turns constructive.
The transition should manifest as a step-like
characteristic in the ring's conductance as a function of the coupling
fields (so far unreported).
However, this reasoning appears to be oversimplified:
the adiabatic condition can not be satisfied in the vicinity of the transition point, since the local steering field
vanishes and reverses direction abruptly at the rim of the ring.
Moreover, typical experimental conditions correspond to moderate field strengths, resulting in nonadiabatic effects in analogy
to the case of spin transport in helical magnetic fields~\cite{betthausen}.
Hence, a more sophisticated approach is required. This includes identifying the role played by nonadiabatic
Aharonov-Anandan (AA) geometric phases \cite{aharonov}.

Here, we report transport simulations showing that a topological phase transition is possible in loop-shaped spin interferometers away from the adiabatic limit. The transition is determined by the topology of the field texture through an effective Berry phase related to the winding parity of the spin eigenmodes around the poles in the Bloch sphere. This contrasts with the actual complexity of the emerging dynamic and AA geometric phases, which exhibit a correlated behavior close to the transition.

We consider a two-dimensional electron gas (2DEG) confined at the interface of a semiconducting heterostructure
($xy$ plane in Fig. \ref{fig-1}a).
The 2DEG is subject to SO interaction due to structure inversion asymmetry, which can be tuned by gate electrodes \cite{nitta}.
The SO field ${\bf B}_{\rm SO}$ couples to conduction electron spin as~\cite{bychkov}
\begin{equation}
H_{\rm SO}=({\alpha}/{\hbar})({\bm \sigma} \times {\bf p})\cdot \hat z \equiv {\bf B}_{\rm SO} \cdot {\bm \sigma},
\label{HSO}
\end{equation}
with ${\bf B}_{\rm SO} = B_{\rm SO} ( {{\hat k} \times {\hat z}})$, $\alpha$ the SO strength, $\bf p$ the electronic momentum,
${\bm \sigma}$ the vector of Pauli spin matrices, ${\hat k}$ the unit vector along the electron wave vector ${\bf k}$,
and $\hat z$ the unit vector perpendicular to the 2DEG. This SO term gives rise to the  Aharonov-Casher (AC) \cite{aharonovcasher} interference
patterns in the conductance of ring ensembles \cite{nagasawa1,nagasawa2}. Geometric and dynamical phases developed by
electrons moving in circular orbits have been identified as distinct contributions to the AC phase in rings~\cite{frustaglia}.
Moreover, spin eigenstates subtend a regular cone in the Bloch sphere with
solid angle $\Omega= - 2 \pi (1-1/\sqrt{Q^2+1})$ where $Q=2m^*\alpha r/\hbar^2$ is the adiabaticity parameter~\cite{frustaglia},
$m^*$ is the effective electron mass and $r$ the ring radius. This corresponds to a geometric AA phase $-\Omega/2$
acquired by the spins in a roundtrip
\cite{frustaglia,nagasawa1}.
The spin states are radial only in the adiabatic limit $Q \gg 1$, giving a Berry phase $\pi$.

We add a homogeneous Zeeman field in the $xy$ plane
\begin{equation}
H_{\rm Z}= {\bf B}\cdot {\bm \sigma}= B (\cos\gamma\, \sigma_x+\sin\gamma\, \sigma_y),
\label{HZ}
\end{equation}
where $\gamma$ is the angle with respect to the axis of the wire. In geometries where the contact leads are symmetrically coupled to the rings, electron
spins traveling along symmetric interference paths acquire equal Zeeman phases resulting in constructive interference for ${\bf B}_{\rm SO} =0$.
Both constructive and destructive interference
of Zeeman phases are possible in rings coupled tangentially to leads to form loops \cite{yang} due to interference of paths shown in Fig.~\ref{fig-1}a.

\begin{figure}
\includegraphics[width=0.95\columnwidth]{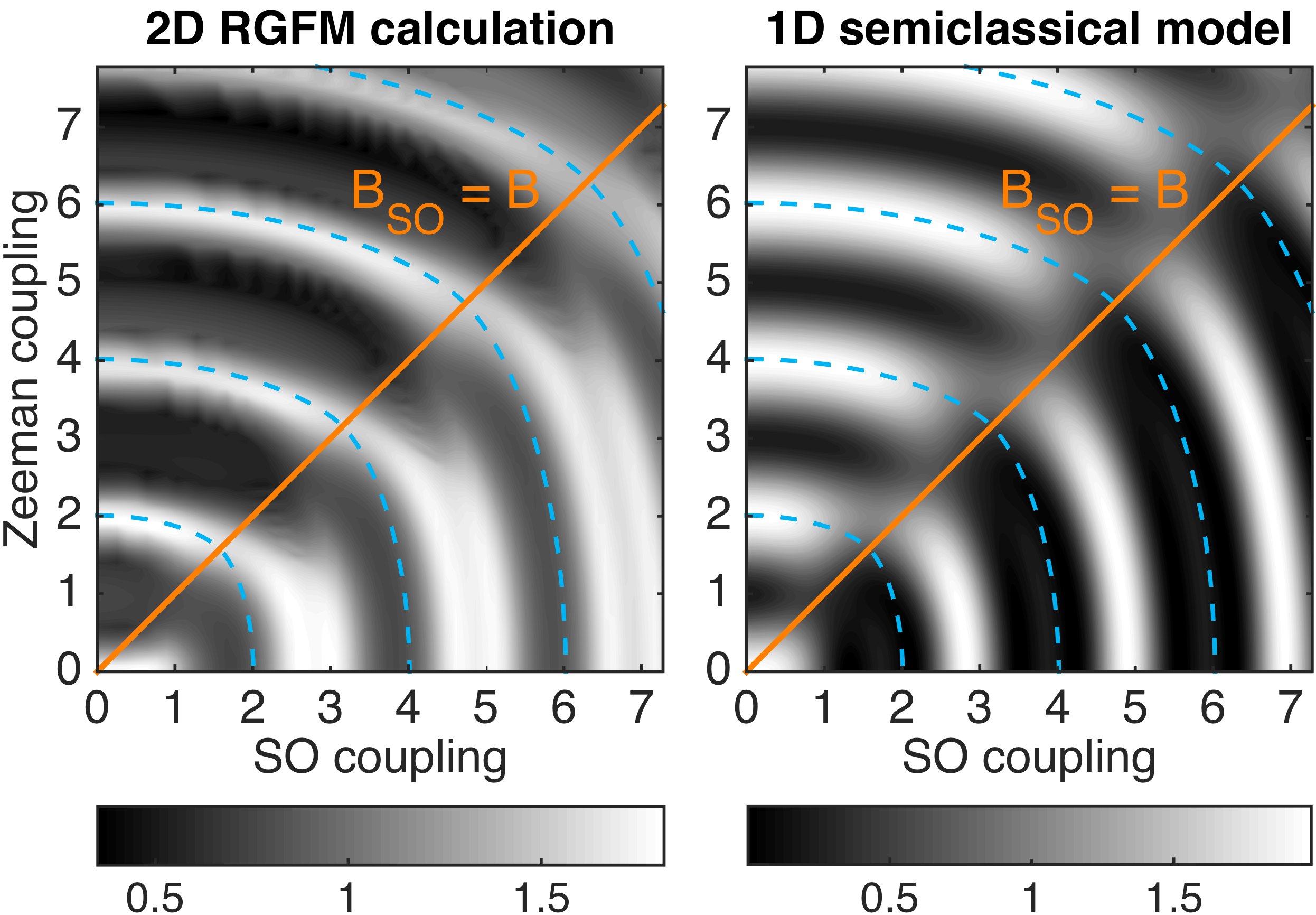}
\caption{
Conductance (in units of $e^2/h$) as a function of the SO and Zeeman couplings in a ballistic single-mode loop.
Left: 2D simulations for a $r=1.2\;{\rm \mu m}$ loop in InGaAs at $E_{\rm F}=88\;{\rm meV}$. Right: 1D semiclassical model.
The dashed lines show the wavefronts in an adiabatic treatment, Eq. (\ref{dynamic}).
The phase dislocation along $B_{\rm SO}=B$ is a signature of transition in field's topology.
The SO and Zeeman scales are in terms of $Q = 2 m^* \alpha r/\hbar^2$ and $2 m^* r B/(\hbar^2k)$, respectively, and $\gamma=\pi/2$.
\label{fig-2}}
\end{figure}
\begin{figure}
\includegraphics[width=\columnwidth]{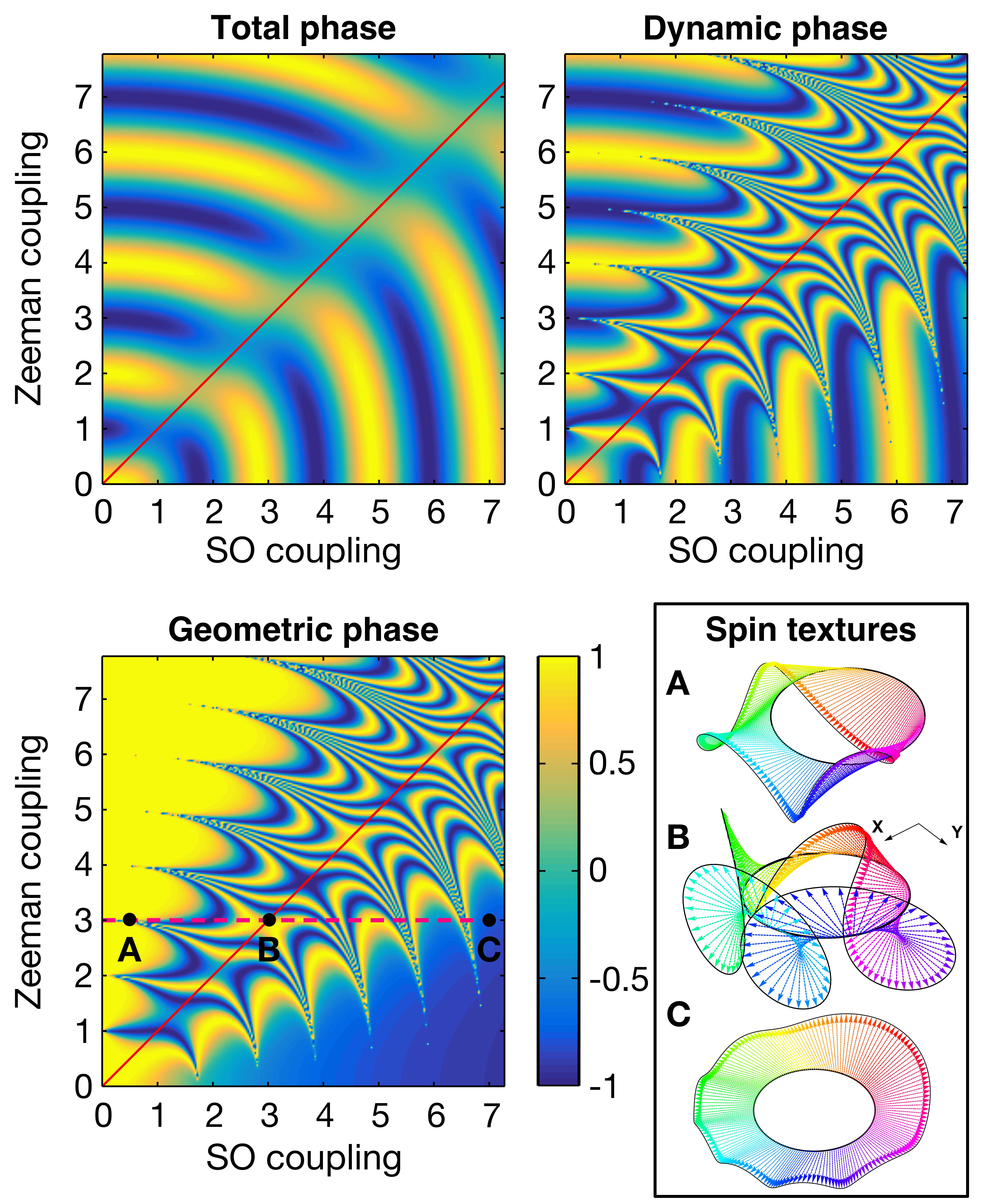}
\caption{Upper panel: cosine of the total phase $\phi$ (left) and the dynamical phase component $\phi_{\rm d}$ (right) 
in the 1D model.
Lower panel: a complementary complexity arises in the cosine of the AA geometric phase component $\phi_{\rm g}$ (left), evidenced by the spin-eigenmode textures calculated at the selected points (right).
\label{fig-3}}
\end{figure}

We adopt here the loop geometry to study the interplay between Zeeman and AC phases.
In the presence of SO coupling the in-plane magnetic field manifests as a pure geometrical effect at the lowest order in $B$,
without affecting the dynamical phase \cite{nagasawa2}.
The perturbation approach fails as $B$ nears $B_{\rm SO}$.
Instead, we use the following methods: i) one-dimensional (1D) calculations based on
semiclassical methods, providing access to local spin dynamics and geometric phases in the ballistic regime, and ii) two-dimensional (2D) numerical simulations
suitable for multi-mode systems with or without disorder.
We assume that the leads are spin-compensated and that the largest energy scale is the Fermi energy $E_{\rm F}$, so that the SO and Zeeman energies
can be considered small in comparison to the kinetic term.
Minor anisotropies arise as a function of $\gamma$ but these are not crucial for our conclusions.

In the 1D semiclassical model we assume three possible and equally probable paths for transmitting spin carriers:
a direct path along the wire and (counter)clockwise paths around the loop (Fig. \ref{fig-1}a).
The $2\times2$ transmission amplitude matrix for spins then reads
$\Gamma \sim \mathds{I}+ \Gamma_+ + \Gamma_-$, where $\Gamma_\pm$ are the (counter)clockwise
transmission amplitude matrices. These are calculated by approximating the circular loop as a regular polygon with a large
number of vertices following the method used in \cite{BFG05}, which is extended here to include in-plane magnetic fields. The conductance is obtained from the transmission probabilities (Landauer formula), given by the trace of $\Gamma \Gamma^\dagger$.

The 2D numerical calculations of electron transport are based on a tight-binding system
of transport equations which was solved using the recursive Green's function method (RGFM)~\cite{wimmer}  as well as the Kwant code~\cite{groth}.
Disorder in the system is introduced by a lattice disorder model 
~\cite{ando}. We use the material parameters of InGaAs ($m^*=0.05m_0$ with $m_0$ the bare electron mass).

\begin{figure}
\includegraphics[width=\columnwidth]{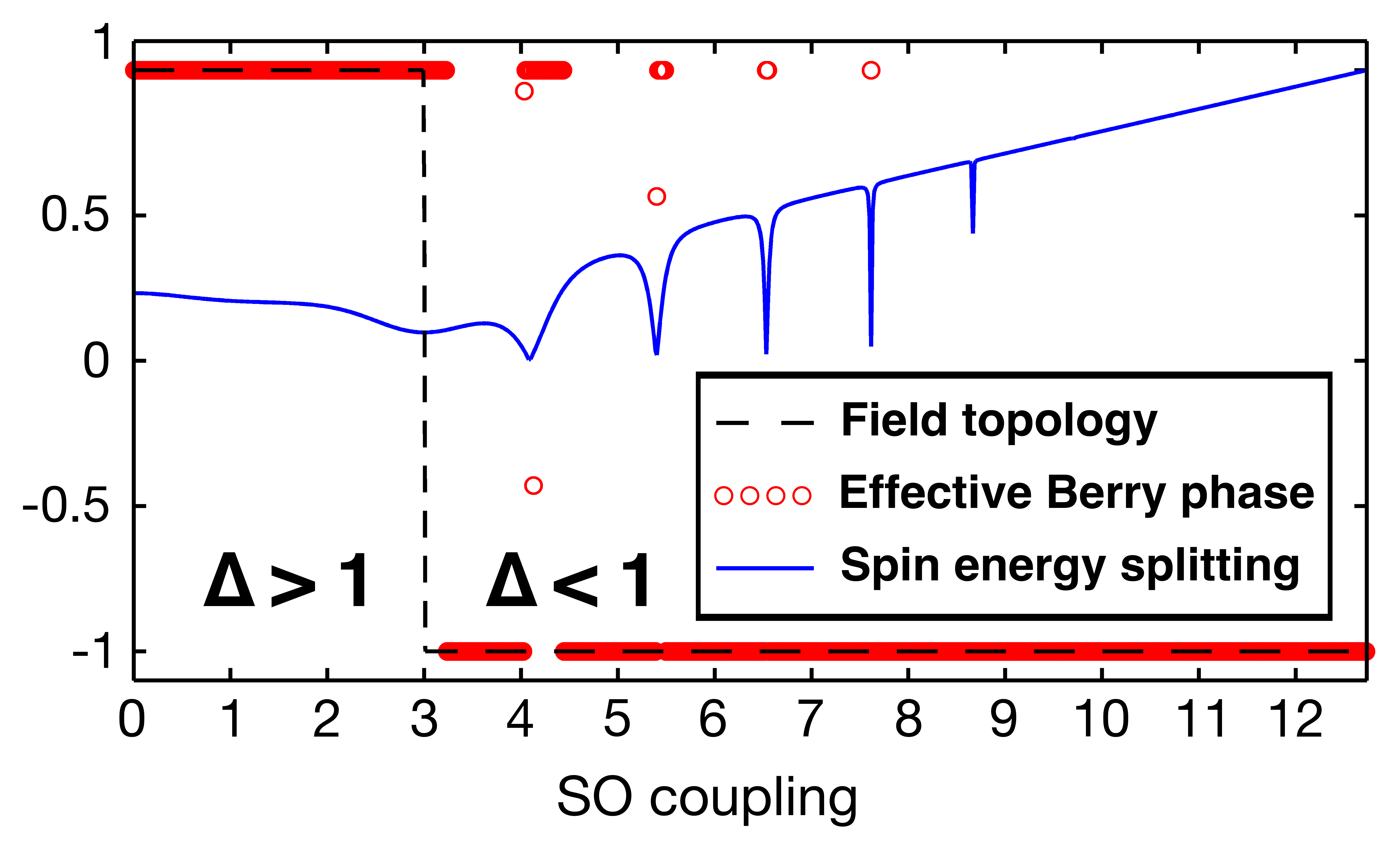}
\caption{Dashed line: cosine of half the solid angle subtended by the
spin-guiding field  along the dashed line in Fig.~\ref{fig-3} (lower panel, left) corresponding to the Berry phase in a hypothetic
adiabatic evolution and subject to a topological transition at $\Delta=1$.
Circles: cosine of the azimuthal component of $\ell \pi$ of the
AA geometric phase $\phi_{\rm g}$ (winding parity) along this path
acting as an effective Berry phase $\phi_{\rm B}$.
Solid line (blue): 1D spin energy splitting between different spin
species (normalized by the largest energy value in that window).
Anomalies arise in $\ell \pi$ near the degeneracy points,
typified by the dips.
\label{fig-4}}
\end{figure}

Figure \ref{fig-2} shows the conductance in a single-mode ballistic loop calculated with both methods.
It displays an interference pattern with two main characteristics:
(i) radial wavefronts starting from the origin and (ii) a distinct phase dislocation along
the critical line $\Delta \equiv B/B_{\rm SO}  = 1$.
The wavefronts correspond to Zeeman oscillations of period $2m^* r B/\hbar^2k=2.0$.
In the adiabatic regime the dynamical spin phase $\phi_{\rm d}$ is proportional to the average field
$\int_0^{2\pi} \sqrt{(B_{\rm SO}\sin \theta + B)^2 + (B_{\rm SO}\cos \theta )^2} \, d\theta$, giving
\begin{eqnarray}
\phi_{\rm d} \propto 2  (B_{\rm SO} + B)\left [( E(\pi/4, {\cal B}  ) + E(3\pi/4, {\cal B} ) \right ],
\label{dynamic}
\end{eqnarray}
where $\theta$ is the angle in Fig. \ref{fig-1}, ${\cal B} = 4B_{\rm SO} B/(B_{\rm SO}+B)^2$, and $E(\varphi,m)$ are elliptic integrals of the 2nd kind.
Lines of constant adiabatic $\phi_{\rm d}$ are plotted in Fig. \ref{fig-2}. 
The fit with the calculated wavefronts is very good despite the fact that actual spin dynamics is nonadiabatic 
(some deviations are visible for $\Delta \ll 1$, where wavefronts are best described by geometric phase shifts \cite{nagasawa2,joibari}).
The critical line corresponds to the frontier where the field texture changes
topology, which coincides with the spin-eigenstate texture only in the adiabatic regime.
These results are intriguing, since the observed pattern presents properties recalling adiabatic dynamics in a nonadiabatic scenario.
The 2D methods give results qualitatively similar to those obtained with the 1D model, indicating that the semiclassical approach captures the essential features.

The main contribution to the 1D results in Fig.~\ref{fig-2} is given by terms of the form
$\Gamma_{\pm}+\Gamma_{\pm}^\dagger$. When diagonalized, these matrices have elements
$\cos \phi_{\pm}^\sigma$ with $\sigma$ the spin-eigenmode label. The phases $\phi_{\pm}^\sigma$ ($\phi$ henceforth)
consist of two parts: $\phi=\phi_{\rm d}+\phi_{\rm g}$, with a dynamical part $\phi_{\rm d}$ and a
geometric AA one $\phi_{\rm g}$. A dimensionless conductance can then be conveniently simplified as
$\mathcal{G} \equiv 1+\cos (\phi_{\rm d}+\phi_{\rm g})$. 
The dynamical spin phase can be obtained independently from the expectation value of the spin
Hamiltonian $H_{\rm s}= H_{\rm SO}+H_{\rm Z}$ over the spin eigenmodes $|\chi(\theta)\rangle$ as
$\phi_{\rm d} = -(m^*r/\hbar^2 k) \int_0^{2\pi} \langle \chi(\theta)|H_{\rm s}|\chi(\theta)\rangle {\rm d}\theta$.
Spin phases $\phi$, $\phi_{\rm d}$ and $\phi_{\rm g}=\phi-\phi_{\rm d}$ together with some typical spin-eigenmode textures are shown in Fig.~\ref{fig-3}.
The phase $\phi_{\rm g}$ behaves smoothly near the axes, approaching the adiabatic limit $\pi$ for a strong radial SO texture and vanishing for
$B_{\rm SO}=0$.
This is apparent from the simple dynamics of the spin eigenstates in those regions (textures A and C). In the vicinity of the critical line $\Delta = 1$, instead, $\phi_{\rm g}$ displays
a complex pattern as a signature of a strongly nonadiabatic spin dynamics (texture B). This shows that an adiabatic treatment~\cite{lyanda-geller} close to $\Delta=1$ is
not suitable even in the limit of strong fields, and no signature of a topological transition is expected in $\phi_{\rm g}$. In contrast, such a transition is indeed present in the total phase $\phi$, visible as a characteristic dislocation in the interference pattern for conductance in Fig.~\ref{fig-2}.

To understand the origin of the topological transition 
we generalize a treatment first introduced in Ref. \onlinecite{ABRPR} for
the study of spin (Berry) adiabatic phases to the case of nonadiabatic spin dynamics.
In the absence of degeneracies, the AA geometric phase can be written as 
$\phi_{\rm g}= \frac{1}{2} \int_0^{2\pi} \frac{\partial\delta}{\partial\theta}(1+\sigma \cos\eta(\theta)) \text{d}\theta = \ell \pi + \frac{\sigma}{2} \int_0^{2\pi} \frac{\partial\delta}{\partial\theta} \cos\eta(\theta) \text{d}\theta$, 
where $\delta$ and $\eta$ are the azimuthal and polar angle coordinates on the Bloch sphere and $\ell$ is an integer
accounting for the windings of the spin eigenmodes around its poles.
The second term in $\phi_{\rm g}$ is responsible for the complex structure shown in Fig.~\ref{fig-3}.
We find that this fluctuating term cancels out exactly with an identical component appearing in
the dynamical phase such that the total phase reduces to $\phi= \phi_{\rm d}^0 + \ell \pi$, where
$\phi_{\rm d}^0=\frac{\sigma}{2} \int_0^{2\pi} \frac{1}{\cos\eta(\theta)} \frac{\partial\delta}{\partial\theta}  \text{d}\theta$
is a smooth component of $\phi_{\rm d}$. Our numerical results show that $\ell$ undergoes
a parity transition near $\Delta=1$, with odd $\ell$ for $\Delta<1$ and even $\ell$ for $\Delta>1$ (Fig.~\ref{fig-4}). Hence, the simplified
dimensionless conductance writes $\mathcal{G}=1+\cos (\phi_{\rm B})\cos (\phi^{0}_{\rm d})$, where we identify $\phi_{\rm B}=\ell\pi$
as an {\it effective} Berry phase causing the phase dislocation at $\Delta=1$ in Fig.~\ref{fig-2} as $\cos (\phi_{\rm B})$ jumps
from $1$ to $-1$, while the smooth term $\phi^{0}_{\rm d}$ leads to wavefronts.
This recalls a topological transition in the adiabatic limit~\cite{lyanda-geller} (dashed line in Fig.~\ref{fig-4}) but
involving an effective Berry phase.

The above picture fails near the degeneracy points \cite{resonances}, where the analyticity of the geometric potentials is not guaranteed. The degeneracy points can be characterized as those for which the dynamical-phase difference between distinct spin species (which is equivalent to the spin energy splitting, Fig.~\ref{fig-4}) is equal to zero. When calculated numerically, $\phi_{\rm B}$ presents a series of anomalies roughly fitting these points. Still, these are compensated by corresponding anomalies arising in $\phi_{\rm d}^0$ such that the total phase $\phi$ is not affected. A full understanding of the role played by degeneracies deserve further efforts beyond the scope of this work.  
Despite that, our approach captures most of the physics relevant to the problem.

\begin{figure}
\includegraphics[width=\columnwidth]{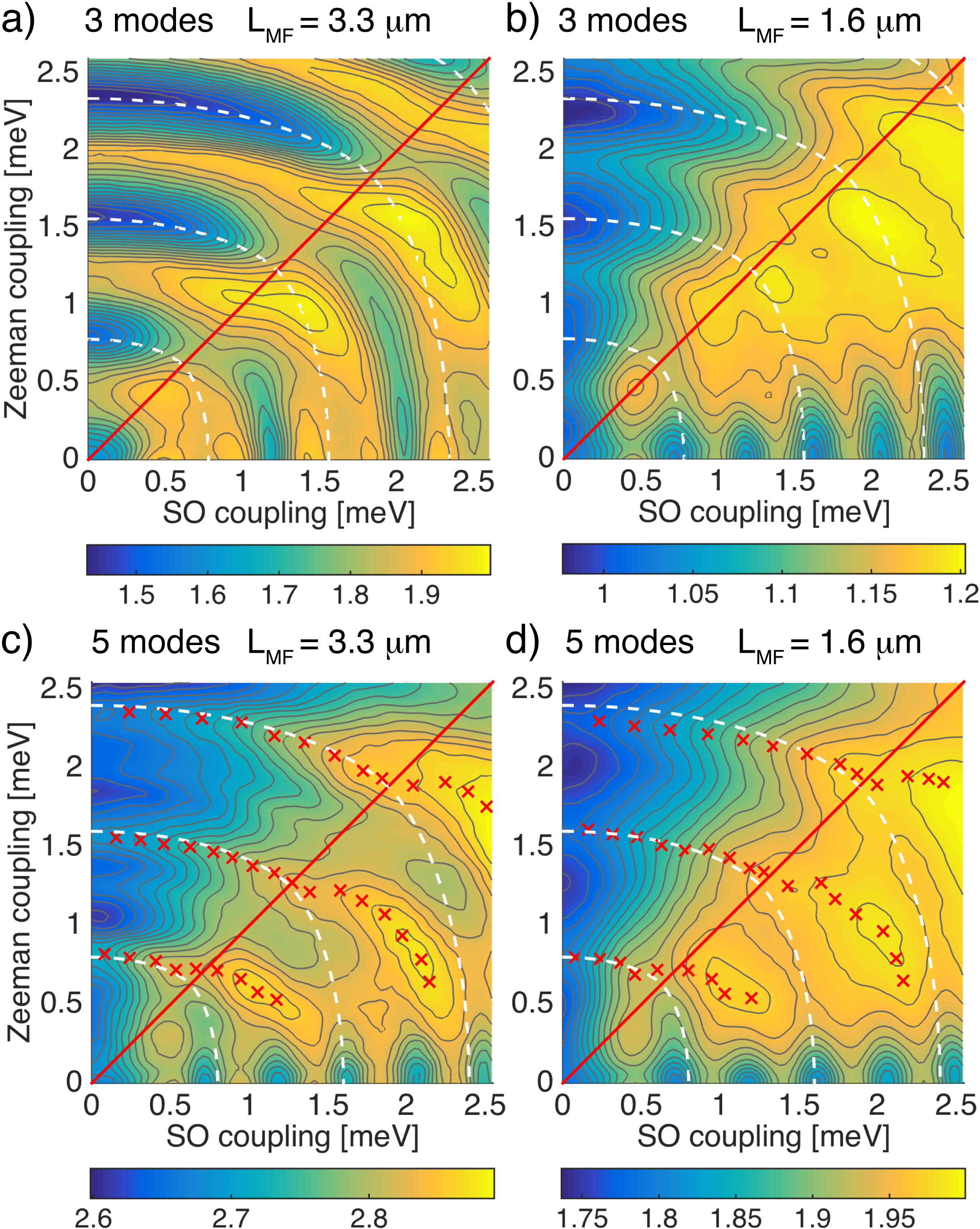}
\caption{Simulated interference pattern in the conductance of multi-mode loops ($r=0.52\;{\rm \mu m}$) calculated with the 2D method.
$E_{\rm F}=64.3\;{\rm meV}$ and $\gamma=0$.
The wire width is $W=35.4\;{\rm nm}$ in (a) and (b), supporting 3 modes, and 53.9 nm in (c) and (d), supporting 5 modes.
Electron mean free path $L_{\rm MF}=3.3\;\mu{\rm m}$ in (a) and (c) and $1.6\;\mu{\rm m}$ in (b) and (d). 
The dashed lines give wavefronts of $\phi_{\rm d}$, Eq. (\ref{dynamic}).
The solid red lines indicate the critical line $\Delta=1$ for the lowest transport mode.
The topological transition is visible as a shift in the interference peak positions for $B_{\rm SO}>B$ (crosses).
\label{fig-5}}
\end{figure}

Experiments are often performed in ensembles of
multi-mode rings where the interference signal is strengthened and nongeneric
features from individual structures are averaged out~\cite{richter}.
Figure \ref{fig-5} shows interference patterns in the conductance of
multi-mode InGaAs loops in the presence of disorder calculated with the RGFM at low temperatures. Zeeman phases are susceptible to temperature and disorder since they are proportional to $1/k$, in contrast to the AC phase which is independent of $k$. Besides, the in-plane field leads to dephasing of the AC oscillations~\cite{meijer}. However, the interference pattern persists in the whole diagram, due to the relevance of Zeeman phases in loops. 
The AC oscillation frequency doubles when the mean free path
decreases as Altshuler-Aronov-Spivak (AAS) paths become relevant~\cite{altshuler}. This effect is not seen for Zeeman phases.
Since $B_{\rm SO}$ is proportional to the propagating velocity of a mode, multiple critical lines may arise. Even though, only the transition of the lowest transport mode is
clearly visible since higher modes move at slower speed, being more prone to scattering and decoherence. 
Nevertheless, the triple-mode case in Fig.~\ref{fig-5}a fits remarkably well the single-mode results for the lowest transport mode~(Fig.~\ref{fig-2}).
These results show that the topological transition is robust, and could be detected in
multichannel loops in the presence of moderate disorder.

We have measured InGaAs samples with mean free paths of the order of a few micrometers~\cite{nagasawa2}.
Analysis of these samples indicates that it is possible to fabricate 0.5 to 1 micron radius loops where the
gate voltage can change $Q$ by about 1.5 to 3 units. A strong 15 T magnetic field gives
$Q$ above 10.  These field ranges are high enough to reveal signatures of the topological transition.
HgTe/HgCdTe is also a good candidate for experiments due to reports showing high mobility~\cite{hgcdte1},
strong $B_{\rm SO}$~\cite{hgcdte2}, and high Zeeman coupling~\cite{KTHSHDSBBM06}.

Our findings open possible lines of future research. Alternative interferometer geometries could be studied with stronger wire-to-ring coupling in comparison to loop geometries allowing for higher signal strength in experiments, e.g., rings with asymmetric interference paths or symmetric rings with Aharonov-Bohm fluxes.
Due to the robustness of the topological transition, a loop device could be used as a magnetometer measuring
the in-situ intensity of the Rashba spin-orbit fields, while deviations from the critical line $\Delta=1$ may be used to estimate the strength of
the Dresselhaus SO interaction \cite{dresselhaus}.
Signatures of complex AA geometric phases may be revealed by studying
transport of spin-polarized carriers \cite{oltscher}.
Finally, we note that analogous topological transitions in geometric phases emerge also in classical physics~\cite{bgoss}.
We have studied magnetic moment dynamics~\cite{landaulifshitz} under
the combined action of rotating and homogeneous fields and found a topological transition
that features a phase shift of $2\pi$ associated with SO(3) rotations.
\begin{acknowledgments}
{\em Acknowledgments.---} 
This work was supported by Grants-in-Aid for Scientific Research (C)
No. 26390014 and (S) No. 22226001 from Japan Society for the Promotion of Science.
DF and JPB acknowledge support from the Spanish Ministry of Science
and Innovation's project FIS2011-29400 and from the Junta de Andaluc\'{\i}a's Excellence Project No. P07-FQM-3037.
\end{acknowledgments}


\end{document}